\documentclass[aps]{revtex4}
\usepackage{graphicx}
\begin{document}
\hfill Applied Physics Report 2004--22

\vspace{0.7cm}

\title{VAN DER WAALS INTERACTION OF PARALLEL POLYMERS AND NANOTUBES}

\author{Jesper Kleis}\thanks{Corresponding author: Tel. +46 31 772 3121;
Fax: +46 31 772 8426; E-mail: \texttt{kleis@fy.chalmers.se}}
\author{Per Hyldgaard}
\author{Elsebeth Schr\"{o}der}
\affiliation{Department of Applied Physics, Chalmers University of Technology and
G\"{o}teborg University, SE--412 96 Gothenburg, Sweden}

\date{July 3, 2004}

\vspace{\parskip}

\begin{abstract}
We study the mutual interactions of simple, parallel polymers and nanotubes,
and develop a scheme to include 
the van der Waals interactions in the framework of density functional theory
(DFT) for these molecules at intermediate to long-range separations. We 
primarily focus on the polymers 
polyethylene, isotactic polypropylene,  
and isotactic polyvinylchloride, but our 
approach applies more generally to all simple polymers and nanotubes.
From first-principle DFT calculations we extract
the electron density of the polymers and their
static electric response. We derive explicit expressions for the
van der Waals interaction energy under simple symmetry assumptions.
\end{abstract}

\maketitle


\vspace{1cm}
\section{Introduction}
Developing an ab initio understanding for 
soft-matter problems, such as found both in the biosciences and 
in carbon-based 
nanotechnology, represents an important challenge. Density functional 
theory (DFT) provides, in principle, the necessary materials-theory 
account but traditional implementations of DFT are insufficient 
for soft matter~\cite{layersurfsc,ijqc}. Traditional DFT 
implementations provide an accurate account of the intra-molecular
(intra-material) binding in regions with a high electron density.
However, the typical soft-matter problem, like the binding of graphitic 
and layered materials~\cite{layerprl,nanovdW2}, of polymer 
matrices~\cite{serra}, and of polyaromatic dimers~\cite{ggpaper,svetlapah},
is defined by regions of sparse electron density. Here traditional
DFT implementations provide no consistent 
account~\cite{serra,scoles,parrinello,mourik}
because they lack a description of the van der Waals (vdW) or 
dispersion forces.  This shortfall of traditional DFT has motivated
a series of development works by us~\cite{ijqc,layerprl,ggpaper} and 
by others~\cite{otherssolve,bostrom,kohnvdW,misquitta} that seeks 
a consistent extension of
traditional DFT (accurate for intra-molecular binding) with
electron-density based calculations of the van der Waals 
interactions (for the intermolecular binding). We thereby extend traditional 
DFT in a consistent approach that avoids
double counting of the electron interaction effects. 

The exchange-correlation energy $E_{xc}[n]$ is a functional
of the electron density $n$ and represents the central concept
in all DFT approximations. It represents the total electron-electron 
interaction energy which is not included in the mean-field 
approximation described by the electron Poisson potential 
$\Phi(\mathbf{r})$.  Together with the external potential 
$V_{\rm ext}$, the functional $E_{xc}[n]$  completely 
determines the general DFT calculational scheme for 
the total energy
\begin{equation}
E = T_0[n] + \int d\mathbf{r} \, n(\mathbf{r}) \,
\left[ \Phi(\mathbf{r})/2+V_{\rm ext}\right] + E_{xc}[n] 
\label{eq:DFTscheme}
\end{equation}
of an interacting soft/sparse or traditional-materials system.
The term $T_0[n]$ in (\ref{eq:DFTscheme}) is the kinetic
energy of a non-interacting system at density $n$ and is in DFT 
straight forward to calculate as a single-particle Schr{\"o}dinger 
equation problem specific to the materials system, as discussed, 
for example, in Refs.~\cite{Harris,DFTbook}.  The functional $E_{xc}[n]$, 
however, is universal and it is this functional that we seek to
improve beyond the traditional local-density and generalized 
gradient approximations in the investigations for a 
new van-der-Waals-density functional (vdW-DF)
\cite{ijqc,layerprl,ggpaper,tractable}. 

In this paper we illustrate the nature of such a new vdW-DF by
ab initio and analytical calculations of the intermediate-range
interactions between 
parallel (unbranched) polymers and nanotubes. Specifically, we 
investigate the interactions between polyethylene (PE), isotactic 
polypropylene (PP), isotactic polyvinylchloride (PVC), and 
single-walled carbon nanotubes (CNT).  We first determine the
structural geometry of the isolated molecules by self-consistent 
DFT calculations using the generalized gradient approximation (GGA),
allowing the atoms to relax to the energetically optimal positions. 
In the optimization the atoms are moved according to the 
Hellmann-Feynman forces in the GGA; details of these calculations 
will be published separately \cite{polyjesper}. 
To this state-of-the-art description 
of the individual molecular structure we here add a charge-density 
based description of the intermolecular binding.

The elongated molecules represent an important class of soft-matter 
interaction problems because they have very anisotropic dielectrical 
response functions, and as such constitute a special problem. 
This interaction is naturally separated in three regimes:
(i) at binding distances where the van der Waals attraction and
kinetic repulsion compete,
(ii) the intermediate regime
where the local electron density variation is important but
where the electron overlap can be neglected, and
(iii) the asymptotic regime
where the van der Waals interaction is given by the overall (as 
opposed to the locally varying) susceptibility
of the molecules and where
traditional 
van der Waals calculations \cite{Unified,classicvdW} become feasible. 
We focus on describing the transition of the molecular van der Waals 
interactions from intermediate separations and out to asymptotic
separations.

The emphasis on intermediate molecule separations $d$ permits us to use a 
simpler approximative treatment of the anisotropic dielectric response, 
which we determine upon averaging the electron density along 
the polymer and nanotube axis. 
Our aim is to find general
trends in polymer-polymer and nanotube-nanotube interactions,
and rather than aiming at numerical, very accurate results
we here develop analytical, approximate expressions for use
in further analyzing the system.
More accurate numerical studies, with improved calculations 
of the molecular dielectrical response, are deferred to a 
forthcoming paper.

\section{The intermolecular vdW binding}
The vdW interactions involve fluctuating (dipole and multipole) 
interactions which
are naturally expressed in terms of the bare (local-field)  and 
external-field susceptibility tensors, $\chi_0$ and $\chi_{\rm eff}$. 
Starting from the adiabatic connection formula \cite{adia1,adia2,adia3} 
the dipole-dipole approximation of the
exchange-correlation energy of two well separated
objects 1 and 2 can be written as \cite{ylvapaper1}
\begin{eqnarray}
E_{xc}^{\mathrm{asymp}}[n]=
E_{\mathrm{vdW}}[n]&\simeq& 
-\int_0^{\infty} \frac{du}{2\pi} 
\int d^3 r_1 \int d^3 r'_1  \int d^3 r_2\int d^3 r'_2
\nonumber\\[0.6em]
&&
\sum_{a,b,c,d}
\chi_{\mathrm{eff},1}^{ab}(\mathbf{r}_1,\mathbf{r}'_1;u) 
T_{12}^{bc}(\mathbf{r}_1,\mathbf{r}_2)
\chi_{\mathrm{eff},2}^{cd}(\mathbf{r}_2,\mathbf{r}'_2;u) 
T_{21}^{da}(\mathbf{r}_2,\mathbf{r}_1)\,.
\label{eq:EvdW}
\end{eqnarray}
Here we allow for \textit{anisotropic\/} effective susceptibility 
tensors of the objects.  
The anisotropy must 
necessarily be taken into account for such extended objects 
as the straight polymers and nanotubes discussed here.
The interaction tensor is given by
$T_{ij}=- \nabla_i\nabla_j |\mathbf{r}_j-\mathbf{r}_i|^{-1}$. 
A standard  frequency integration has been turned into an integral
on the imaginary axis, with $u=- i\omega$ and $\omega$ 
the physical frequency.
Hartree atomic units are used throughout, unless otherwise noted.

In this study we seek only to explore the formal nature of the
interactions between parallel polymers and nanotubes.
We therefore make use of a simplified but electron-density 
based model description of the bare molecular susceptibility
$\chi_0$ which is assumed diagonal. This model assumption implies 
that the anisotropic dielectric response $\chi_{\rm eff}$ (that 
characterizes the external-field response of polymers/nanotubes) is
taken to arrive exclusively by the local-field effects that are
specified by this diagonal $\chi_0$.

For the diagonal elements of $\chi_0$ we use the approach of
Refs.~\cite{nanovdW2,nanovdW1}, that is, a slightly modified version 
of the effective susceptibility used in Ref.~\cite{Unified} 
\begin{equation}
\chi_0(\mathbf{r},\mathbf{r'};u)=\delta(\mathbf{r}-\mathbf{r'})
\frac{n(\mathbf{r})}{u^2+u_0^2}.
\label{eq:locsusc}
\end{equation}
The valence electron density $n(\mathbf{r})$ is obtained directly
from the above-mentioned self-consistent DFT calculations. 
The cut-off frequency $u_0$ was introduced in Eq.~(\ref{eq:locsusc})
to avoid an unphysical divergence in the static limit ($u=0$).
The cut-off
is determined from a separate
set of self-consistent DFT calculations that provide results for the
static polarization by imposing a finite external electrical field 
\cite{nanovdW2,nanovdW1} perpendicular to the polymer and nanotube 
axis.

\begin{figure}
\begin{center}
\scalebox{0.7}{\includegraphics{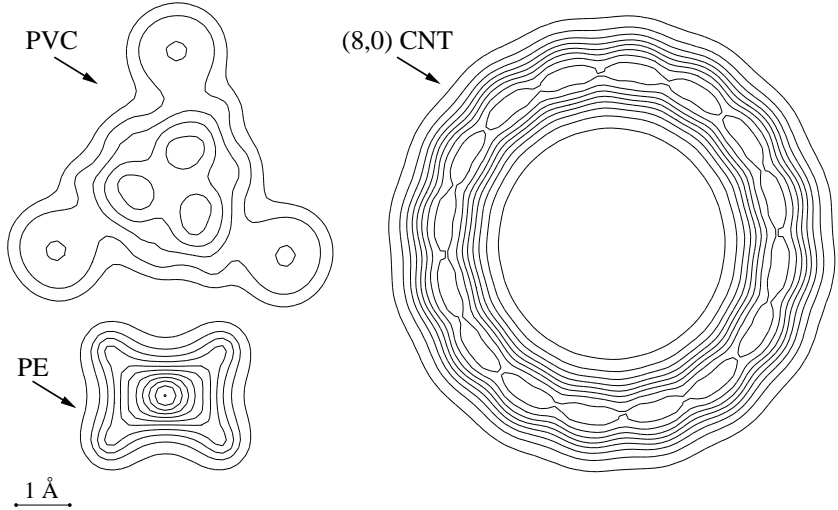}}
\caption{
Valence charge density $n(\mathbf{s})$ of polyethylene (PE), polyvinylchloride (PVC), 
and the $R_{(8,0)}$-nanotube (CNT), averaged along the length of the molecules. 
Contour line spacing is 0.15 e{\AA}$^{-3}$ and only densities equal to or
  above this value are shown. The valence charge density plot of polypropylene (PP)
looks qualitatively  similar to that of PVC.
\label{figdens}} 
\end{center}
\end{figure}

Figure~\ref{figdens} shows the electron density (averaged along
the polymer and nanotube axis) as calculated by traditional DFT
for the molecules PE, PVC, and CNT. The averaging along the length
is done in the
same spirit as our calculations for the vdW interactions in
layered materials~\cite{ijqc,layerprl,tractable}. The averaging
provides a qualitative picture of the electron-density 
variation that effectively determines the intermolecular
vdW interaction at the relevant intermediate molecular
separations. For simplicity  we choose to base the set of
numerical and analytical evaluations of the vdW
correction (\ref{eq:EvdW}) on dielectric-response functions
specified by these averaged electron densities.

\begin{figure}
\begin{center}
\scalebox{0.7}{\includegraphics{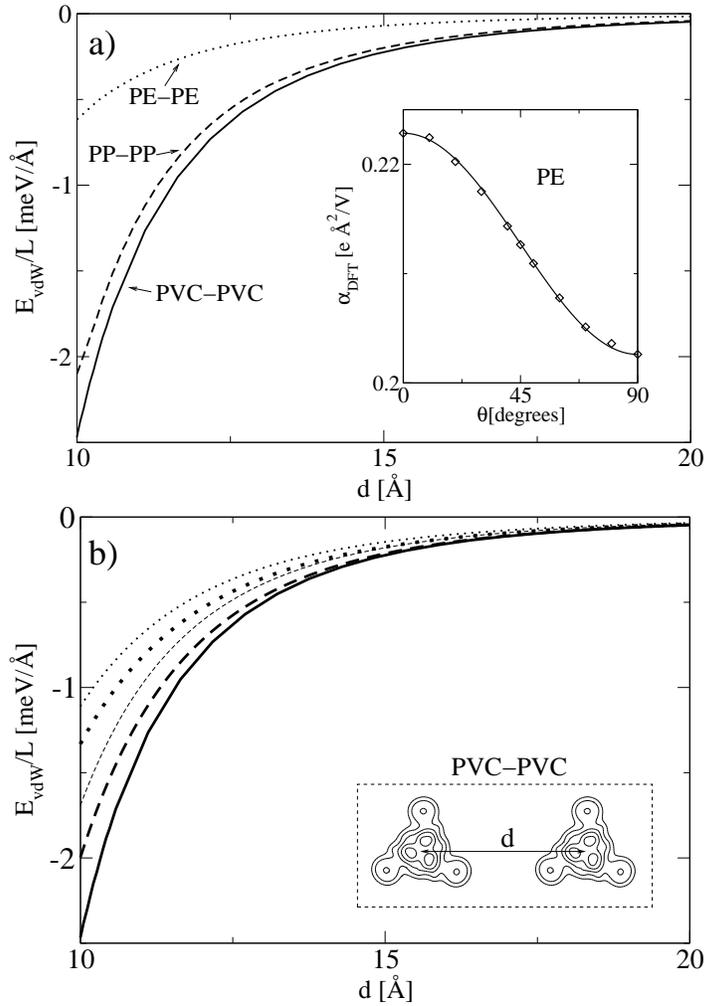}}
\caption{The van der Waals part of the interaction energy for two parallel
polymers at separation $d$. 
Panel (a) shows 
$E_\mathrm{vdW}/L$ numerically evaluated for sets of length-averaged polymers
keeping their full angular structure. 
The squares in the inset show the DFT calculated susceptibility of PE for
different angles. The full line is a fit based on the rotational properties
of the susceptibility tensor.  
Panel
(b) shows the results for the vdW interaction energy of PVC in 
 different approximations. 
The thick lines correspond to evaluations keeping the full angular structure
for the length-averaged electron density, and the thin
lines correspond to evaluations of the radially averaged density. 
Shown are expansions to 5th order in $d^{-1}$ (dotted lines), 
to 7th order (dashed lines), and the full calculation without expansion
 (solid line). 
In both panel (a) and (b) the PVC molecules 
are oriented as shown for separation $d=10$ {\AA} in the inset of (b). 
The PP molecules are oriented
similarly to PVC, and the PE molecules are oriented with the long 
axis towards each other.
\label{figenergy} 
}
\end{center}
\end{figure}

Figure~\ref{figdens} also suggests that this effective electron
distribution has an approximate cylindrical rotational symmetry which
we utilize for further simplifications in our present model 
study. We term such molecules `cylindrical' and these
remain the focus of this paper because they offer in several
cases comparison against analytical expansions and
evaluations.  We stress that for the question of the 
mutual vdW interactions the most relevant measure of cylindrical
symmetry is exactly the angle-dependence of the static external
field response.  In the set of 
investigated molecules this angle-dependence is small,
as shown for PE in the inset of Figure \ref{figenergy}(a)
from direct evaluation in traditional DFT.

An element of charge at position $\mathbf{s}$ (specified 
in two dimensions relative
to the molecule center-of-mass line) on the molecule responds to the 
local electric field according to $\chi_0(\mathbf{s};u)$ given by (\ref{eq:locsusc}).
The local field is the applied electric field 
$\mathbf{E}_{\mathrm{applied}}(u)$ (which arises from charge fluctuations
at the other molecule) screened by the presence 
of other elements of charge in the neighborhood.
The local electric field is described via the local electric potential
$\phi(\mathbf{s};u)$ and given by charge conservation
\begin{equation}
-\nabla \cdot\left\{ \left(  1+4\pi\chi_0\right)
\nabla \phi  \right\}=0\,\,.
\label{eq:chargecon}
\end{equation}
The effective susceptibility tensor $\chi_{\mathrm{eff}}$,
which appears in the expression for the interaction energy (\ref{eq:EvdW}),
is related to the local susceptibility and the local electric potential by
\begin{equation}
\chi_{\mathrm{eff}}(\mathbf{s};u) \mathbf{E}_{\mathrm{applied}}(u) = 
-\chi_0(\mathbf{s};u)\nabla\phi(\mathbf{s};u)\,.
\label{eq:chieff}
\end{equation}
Due to the assumed translational invariance along the length of the
molecule we can in general solve (\ref{eq:chargecon}) numerically as 
a two-dimensional differential equation in $\phi(\mathbf{s};u)$ with
the frequency $u$ as a parameter, and thus find 
$\chi_{\mathrm{eff}}(\mathbf{s};u)$ from (\ref{eq:chieff}).
The spatial integral 
$\alpha(u)=L \int d^2 s \chi_{\mathrm{eff}}(\mathbf{s};u)$ 
taken at $u=0$ gives us the 
molecular, macroscopic susceptibility tensor as a function of $u_0$.
Comparing $\alpha(u=0)$ to the static
DFT-calculated molecular susceptibility we determine the frequency 
constant $u_0$.

In polymers our approach \cite{nanovdW2,nanovdW1}
yields a value of the static polarization 
which varies with the angle at which the 
electric field is applied. However, for the simple polymers
the variation is small, as shown for PE in the inset of Figure~\ref{figenergy}(a).
For PP and PVC the susceptibility is
almost constant with respect to rotation, with 
$\alpha^\mathrm{PP}=0.97\pm0.01$ e\AA$^2$/V
and $\alpha^\mathrm{PVC}=1.01\pm0.01$ e\AA$^2$/V. 
The values $u_0$ found 
for the angular means of the static polarizability
 are \cite{polyjesper} $0.429$ Ha 
for PE, $0.507$ Ha for PP, and $0.482$ Ha for PVC.
In nanotubes $u_0$ depends on 
the nanotube radius. However, the variation is 
small and the macroscopic susceptibility is not very sensitive 
to this variation. For CNT we thus use 
the large-nanotube value \cite{nanovdW1} for all nanotubes, $u_0=0.30$ Ha.

In the expression for the van der Waals energy (\ref{eq:EvdW}) the two 
spatial integrals along the (density averaged) molecules can be
performed analytically \cite{nanovdW2}, but leave a cumbersome integral 
over the remaining spatial variables and the frequency.
These integrals can be numerically solved, and the results
for pairs of parallel polymers are shown in Figure~\ref{figenergy}(a)
for a specific relative orientation of the polymers. The results
for other relative orientations will be reported in Ref.~\cite{polyjesper}.

\section{Interaction energy in cylindrical symmetry}

Whereas it is possible to numerically evaluate $E_{\mathrm{vdW}}$
for a general pair of parallel molecules, 
such results provide only limited insight into the vdW interaction 
mechanism. To interpret 
such numerical evaluations we provide instead analytical
results that are possible to derive with our choice of
simplified local susceptibility function (\ref{eq:locsusc})
for cylindrical molecules.
When cylindrical symmetry is imposed, the local electric potential
$\phi$ factorizes into 
$\phi(\mathbf{s};u)=-|\mathbf{E}_{\mathrm{applied}}(u)|W(s;u)\cos\theta$,
where $\theta$ measures the angle from the direction of the applied field.
Solving (\ref{eq:chargecon}) then reduces to 
the one-dimensional problem of finding $W(s;u)$ with $u$ as
a parameter. In Ref.~\cite{polyjesper} the more complex problem of
non-cylindrical polymers will be treated in greater detail.

For cylindrical molecules
$\chi_{\mathrm{eff}}$ is
diagonal in the cylindrical representation $(s,\theta,z)$ \cite{nanovdW1},
and the interaction energy reduces to
\begin{equation}
\frac{E_{\mathrm{vdW}}}{L}=
-\int_0^\infty \frac{du}{2\pi} \int_0^\infty d s_1 s_1 \int_0^\infty
d s_2 s_2 \sum_{\beta,\gamma=s,\theta,z}
\chi_\mathrm{eff}^{\beta}(s_1;u) \chi_\mathrm{eff}^{\gamma}(s_2;u) 
G_{\beta\gamma}(s_1,s_2)
\label{eq5}
\end{equation}
where the geometry factors are
\begin{equation}
G_{\beta\gamma}(s_1,s_2)=
\int_0^{2\pi} d \theta_1\int_0^{2\pi} d \theta_2 
\int_{-\infty}^\infty d (z_2-z_1)
\left[
T^{\beta\gamma}(s_1,\theta_1,z_1,s_2,\theta_2,z_2) 
\right]^2\,\,.
\label{eq6}
\end{equation}
The integral over $z_2-z_1$ \cite{nanovdW2} and one of the
angles in $G_{\beta\gamma}(s_1,s_2)$
can 
be carried out analytically, but still leads to intractable
expressions in the remaining angle,
$s_1$, $s_2$, and $d$, and the full presentation is deferred.

Our further treatment of the interaction energy expression
differs depending on the nature of the molecules considered.
On the one hand the polymers considered here are thin 
(radius $ \approx 2 $\ {\AA}) compared to the typical nanotube 
(radius $ \approx 10$\ {\AA}) and compared to the typical 
binding separation ($ \approx 3$--4 {\AA} plus radii of molecules).
This means that a large-separation approximation, based on the
center-of-mass to center-of-mass distance $d$, is a better 
approximation for the polymers than for nanotubes. 
On the other hand, nanotubes are about the most cylindrically
symmetric extended molecules that exist, and the major
part of the electron density is concentrated in a thin 
cylindrical shell, as shown in Figure~\ref{figdens}. 
It is therefore reasonable to approximate the nanotube electron charge
density by a weighted radial delta-function. We further 
approximate the nanotube effective susceptibility tensor by assuming
$\chi_{\mathrm{eff}}^s = \chi_{\mathrm{eff}}^\theta$.
Below, we pursue these approximations for polymers and for
nanotubes.  We also show that the two sets of derivations
are consistent under relevant conditions.

\section{Polymer interaction}

As argued above, we can explore the polymer interactions in an expansion
in powers of $d^{-1}$. Expanding the geometry factors, 
writing $G_{\beta\gamma}=\sum_i G_{\beta\gamma}^{(i)} d^{-5-i}$, 
the first three even-index terms are 
\begin{eqnarray}
G^{(0)}&=&
\frac{9\pi^3}{128}
N
\label{eq:iabexpansion}
\\[0.6em]
G^{(2)}(s_1,s_2)&=&
\frac{225\pi^3}{1024}
\left((N+2M) s_1^2 + 
(N+2M)^T s_2^2
\right)
\label{eq:iabexpansion2}
\\[0.6em]
G^{(4)}(s_1,s_2)&=&
\frac{3675 \pi^3}{8192}
\left((N+4M) s_1^4 + 3 L s_1^2 s_2^2 + (N+4M)^T s_2^4 \right)
\label{eq:iabexpansion4}
\end{eqnarray}
with the tensors
\begin{equation}
N=\left(\begin{array}{rrr}
19 & 19 & 10\\
19 & 19 & 10\\
10 & 10 & 12\\
\end{array}
\right), 
\hspace{1em}
M=\left(\begin{array}{ccc}
12 & 12 & \,8\, \\
7 & 7 & 2\\
5 & 5 & 6\\
\end{array}
\right),
\hspace{1em}
L=\left(\begin{array}{rrr}
99 & 73 & 52\\
73 & 59 & 28\\
52 & 28 & 48\\
\end{array}
\right).
\end{equation}

For cylindrical molecules all terms odd in $i$ for symmetry reasons vanish 
after integration in (\ref{eq5}),
the odd-index terms being a measure of the molecular charge anisotropy.
The polymers treated here in reality are cylindrical symmetric
to a varying degree.
We find that for PE interactions 
the numerical values of the first few odd-index contributions 
to the energy are indeed small \cite{polyjesper}.

To lowest order in $d^{-1}$ the interaction energy becomes
\begin{eqnarray}
\frac{E_{\mathrm{vdW}}^{(0)}}{L}
&=& -
\sum_{\beta,\gamma=s,\theta,z} 
G_{\beta\gamma}^{(0)}d^{-5}
\int_0^\infty \frac{du}{2\pi} 
\left(\int_0^\infty d s_1 s_1\chi_\mathrm{eff}^{\beta}(s_1;u) \right)
\left(\int_0^\infty d s_2 s_2\chi_\mathrm{eff}^{\gamma}(s_2;u)  \right)
\nonumber \\[0.6em]
&=&
-\frac{9\pi^2}{256\,d^5}
\sum_{\beta,\gamma=s,\theta,z} 
N_{\beta\gamma} J_{\beta\gamma}
\label{NJ}
\end{eqnarray}
where we have defined the frequency integral
\begin{equation}
J_{\beta\gamma}=
\int_0^{\infty} du
\overline{\chi_{\mathrm{eff},1}^{\beta}}(u)\,
\overline{\chi_{\mathrm{eff},2}^{\gamma}}(u)\,
\label{eq:Jdef}
\end{equation}
given by the
spatial integral of $\chi_{\mathrm{eff}}^{\beta}$
\begin{equation}
\overline{\chi_{\mathrm{eff}}^{\beta}}(u)
=\frac{1}{2\pi L}\int d^3 r \chi_{\mathrm{eff}}^{\beta}(s;u) 
= \int_0^\infty ds s \chi_{\mathrm{eff}}^{\beta}(s;u) \,.
\label{eq:chibardeff}
\end{equation} 
In the approximation $\overline{\chi_{\mathrm{eff}}^{s}}(u)=
\overline{\chi_{\mathrm{eff}}^{\theta}}(u)$, used below for 
nanotubes, we find from (\ref{NJ}) the asymptotic contribution to 
the interaction energy
\begin{equation}
\frac{E_{\mathrm{vdW}}^{(0)}}{L}=
-\frac{9\pi^2}{256d^5}
\left(4\cdot 19 J_{ss}+2\cdot 10J_{sz}+2\cdot 10J_{zs}+12 J_{zz} \right)
\label{eq:Jssetc}
\end{equation}
Using (\ref{eq:iabexpansion2}) and (\ref{eq:iabexpansion4}) and the
corresponding generalization of (\ref{eq:Jdef}) and (\ref{eq:chibardeff}) 
the next order corrections can be similarly expressed.

In Figure~\ref{figenergy}(b) the $d^{-1}$-expansion is illustrated
along with a numerical calculation that keeps the full angular dependence 
not assuming any rotational invariance (solid line). 
The energy is for the PVC-PVC interaction
in the relative arrangement shown in the inset. This is the 
 orientation affected the most by the anisotropy
of the polymer. The thin dotted line is the lowest order contribution
(\ref{NJ}) of the expansion. As the next non-vanishing order, $d^{-7}$, is added
the description is improved (thin dashed line). 
We checked that adding the 9th order did
improve the description of the energy slightly, but not 
significantly at the separations
considered here. As a comparison, we also show the $d^{-1}$-expansion
of the full angular calculation: the thick dotted line shows the $d^{-5}$
term, and the thick dashed line includes all orders up to  $d^{-7}$.
 We note that using the assumption of cylindrical
symmetry of PVC does worsen the quality of the evaluation somewhat, but already 
including a few terms in the $d^{-1}$-expansion improves the approximations.

\section{Nanotube interaction energy}

For nanotubes a useful simplification of (\ref{eq5}) and (\ref{eq6}) 
can be obtained by a different 
approximation. Rather than using the  $d^{-1}$-expansion,
we stress  that the nanotube electron density
is relatively well described as a weighted radial delta-function
located on the nanotube radius $R$ \cite{nanovdW1}. This leads to
the approximation 
$\chi_{\mathrm{eff}}^\beta(s;u)
\approx 
\overline{\chi_{\mathrm{eff}}^{\beta}}(u)\delta(s-R)/R
$, which decouples the susceptibility factors from
the spatial integration, and we find 
\begin{eqnarray}
\frac{E_{\mathrm{vdW}}}{L}&=&
-\sum_{\beta,\gamma=s,\theta,z}
\int_0^\infty \frac{du}{2\pi}\,
\overline{\chi_{\mathrm{eff},1}^{\beta}}(u)\,
\overline{\chi_{\mathrm{eff},2}^{\gamma}}(u)\,
G_{\beta\gamma}(R_1,R_2)
\nonumber \\[0.6em]
&=& -\frac{1}{2\pi} \sum_{\beta,\gamma=s,\theta,z} 
J_{\beta\gamma}\, G_{\beta\gamma}(R_1,R_2).
\label{Evdw2}
\end{eqnarray}
This is the expression for the interaction of two parallel
nanotubes of radii $R_1$ and $R_2$. 
The integrals in $G_{\beta\gamma}(R_1,R_2)$ yield, in general,
tedious expressions with one remaining angular integral.
However, with the introduction of another, numerically
small approximation of the effective susceptibility tensor
we can carry the analysis of $E_{\mathrm{vdW}}$ further
with simple expressions. 

Assuming that 
the averaged effective radial and tangential susceptibilities 
on each molecule are identical, $\overline{\chi_{\mathrm{eff}}^{s}}(u)=
\overline{\chi_{\mathrm{eff}}^{\theta}}(u)$, then 
$J_{ss}=J_{s\theta}=J_{\theta s}=J_{\theta\theta}$,
$J_{sz}=J_{\theta z}$, and $J_{zs}=J_{z\theta}$.
This is a less restrictive approximation than
the already modest approximation $ \chi_{\mathrm{eff}}^{s}(s;u)=
\chi_{\mathrm{eff}}^{\theta}(s;u)$ imposed 
in Refs.~\cite{nanovdW2,nanovdW1}.
The van der Waals energy per nanotube length then becomes~\cite{nanovdW2}
\begin{eqnarray}
\frac{E_{\mathrm{vdW}}}{L}
&=& -\frac{1}{2\pi}\Big\{
J_{ss}\Big( G_{ss}+G_{s\theta}+G_{\theta s}
+ G_{\theta \theta}\Big)
+ J_{sz} \Big( G_{sz}+G_{\theta z} \Big)
\nonumber\\[0.6em]
&& {}
+ J_{zs} \Big( G_{zs}+G_{z \theta} \Big)
+ J_{zz} G_{zz}\Big\}
\label{eqeji}
\end{eqnarray}
where $G_{\beta\gamma}$ is implicitly taken in nanotube radii $R_1$ and $R_2$.

In Eq.~(\ref{eqeji}) all the sums of $G$-terms are of the same form,
$G_{ss}+G_{s\theta}+G_{\theta s} + G_{\theta \theta}=19 G_{\mathrm{tot}}$,
$G_{sz}+G_{\theta z}=G_{zs}+G_{z \theta}=5 G_{\mathrm{tot}}$ and
$G_{zz}=3G_{\mathrm{tot}}$ where 
\begin{eqnarray}
G_{\mathrm{tot}}(R_1,R_2)&=&
\frac{9\pi}{128}\int_0^{2\pi} d \theta_{1}\int_0^{2\pi} d \theta_{2}
\left((R_2\cos\theta_2+d-R_1\cos\theta_1)^2+
(R_2\sin\theta_2-R_1\sin\theta_1)^2\right)^{-5/2}
\nonumber \\[0.6em]
&=&
\frac{9\pi^2}{16\,d^5}\,\,
\int_{0}^{\pi/2} d\xi
\,\,{}_2F_1\left(5/2,5/2;1;
\{(R_2-R_1)^2+4R_1R_2\sin^2\xi\}/d^2\right)\,.
\label{eq:nanoGtotR1R2}
\end{eqnarray}
The second line is a significant simplification which expresses
the geometry variation in terms of the hypergeometric function
${}_pF_q$ integrated over the variable 
$\xi =(\theta_2-\theta_1)/2$.

For $R_1=R_2=R$ also the remaining integral can be carried out,
again in terms of 
a hypergeometric function \cite{glasser}. 
We can thus factorize $E_{\mathrm{vdW}}$ into one frequency integral 
$J_{\mathrm{tot}}= 19J_{ss}+5J_{zs}+5J_{sz}+3J_{zz}$
and one spatial integral $G_{\mathrm{tot}}(R,R)$
\begin{eqnarray}
\frac{E_{\mathrm{vdW}}}{L}&=&
-\frac{1}{2\pi}J_{\mathrm{tot}}\,G_{\mathrm{tot}}(R,R)
\nonumber\\[0.6em]
&=&-\frac{9\pi^2J_{\mathrm{tot}}}{64\,d^{5}}
\,\,
{}_3F_2\left(1/2,5/2,5/2;1,1;4 R^2/d^2\right)\,.
\label{eq:3F2}
\end{eqnarray}
The only numerical integration needed is the frequency integral
in $J_{\mathrm{tot}}$.
This simplification previously allowed us to extract the graphite-graphite
limit of two large-radius nanotubes at intermediate separation 
\cite{nanovdW2}. It also allows us to interpret the above-mentioned
$d^{-1}$-expansion of cylindrical molecules:
By expanding (\ref{eq:3F2}) to the lowest non-vanishing order in $d^{-1}$
we recover (\ref{eq:Jssetc}).

\section{Conclusions}
We have here presented a general approach based on first-principles electron density
calculations for computing the intermediate to long-range interactions of
parallel, geometrically simple polymers and nanotubes. 
We have applied the scheme to polyethylene, isotactic polypropylene,  
and isotactic polyvinylchloride, and the interaction energy for these 
molecules has been evaluated. We argue that the polymer electron charge density 
 can be approximately described as rotationally invariant,
which allows us to derive a number of simple explicit expressions
for qualitative insight. For parallel nanotubes we derive expressions
for the interactions in terms of hypergeometric functions.
Further, the asymptotic expansion for the nanotubes
is seen to be consistent with the
expression obtained for the cylindrical polymers. 

\section{Acknowledgments}

We thank L.~Glasser for helpful comments and for suggesting the
equal-radius special-function evaluation (\ref{eq:3F2}).
This work was partly supported by
the Swedish Research Council (VR),
the Swedish National Graduate School in Materials Science,
the EU Human potential research training network 
ATOMCAD under contract number HPRN-CT-1999-00048, and
the Swedish Foundation
for Strategic Research (SSF) through the consortium ATOMICS.


\end{document}